\documentclass[]{article}
\usepackage{amssymb}
\usepackage{amsmath}
\usepackage{graphicx}
\usepackage{subcaption} 
\usepackage{cases}
\usepackage{epstopdf}
\usepackage{epstopdf}
\usepackage{bbm}
\usepackage{cite}
\usepackage{stmaryrd} 
\usepackage{empheq}
\usepackage{float}
\usepackage{footmisc}
\usepackage[makeroom]{cancel}
\usepackage[table,xcdraw]{xcolor}
\usepackage{hyperref}
\usepackage{enumitem}
\usepackage{mathtools}
\usepackage{placeins}
\usepackage{textcomp}
\hypersetup{
	colorlinks=true,
	linkcolor=blue,
	filecolor=magenta,      
	urlcolor=blue,
}

\usepackage{stmaryrd}
\usepackage{titlesec} 

\usepackage{geometry}
\renewcommand\thesection{\Roman{section}} 
\renewcommand\thesubsection{\roman{subsection}} 
\titleformat{\section}[block]{\large\scshape\centering}{\thesection.}{1em}{} 
\titleformat{\subsection}[block]{\large}{\thesubsection.}{1em}{} 

\title{G3M Impermanent Loss Dynamics}
\author{Nassib Boueri\footnote{\href{mailto:nassib@cadmos.finance}{nassib@cadmos.finance} } }

\date{September 10, 2021}

\begin{document}
	
	\maketitle
	
	\begin{abstract}
		Geometric Mean Market Makers (G3M) such as Uniswap, Sushiswap or Balancer are key building blocks of the nascent Decentralised Finance system. We establish non-arbitrage bounds for the wealth process of such Automated Market Makers in the presence of transaction fees and highlight the dynamic of their so-called \textit{Impermanent Losses}, which are incurred due to negative convexity and essentially void the benefits of portfolio diversification within G3Ms.
		
		We then turn to empirical data to establish if transaction fee income has historically been high enough to offset Impermanent Losses and allow G3M investments to outperform their continually rebalanced constant-mix portfolio counterparts. It appears that the median liquidity pool had a net nil ROI when taking Impermanent Losses into account. The cross-sectional dispersion of ROI has however been high and the pool net ROI ranking has been significantly autocorrelated for several weeks. This suggests that G3M pools are not yet efficiently arbitraged as agents may access ex-ante knowledge of which G3M pools are likely to be far better investment proposals than others.
		
		We finally focus on the UniswapV3 protocol, which introduced the notion of concentrated liquidity ranges and show that such a position can be replicated by leveraging a classic UniswapV2 pool while simultaneously hedging part of the underlying token price exposition. As such, the herein described Impermanent Loss dynamics also apply to UniswapV3 pools.
	\end{abstract}

	\section{Set-up}
	
	We consider a n-dimensional vector of risky assets $S := (S_1,\dots,S_n)$ of dynamic:
	
	\begin{equation}
		S(t=0) = (1,\dots,1),\frac{dS}{S} = \mu_t dt +  dW_t
	\end{equation}
	
	\noindent
	with $\mu$ a smooth functional of time and $W$ a n-dimensional Brownian motion of smooth covariance matrix process:  $$\Sigma := (\rho_{i,j}\sigma_i\sigma_j)_{t\ge0}.$$
	
	\bigskip
	
	\noindent

	An Automated Market Maker (AMM) is an investment strategy which consists in locking a portfolio of risky assets within a pool (called \textit{liquidity pool}) and to automatically make a market between the pool components. When the price at which a market is made only depends on the current portfolio holdings, the AMM is said to be a Constant Function Market Maker (CFMM). In other words, third party agents may freely add and withdraw assets from the pool as long as a given functional of the pool composition remains unchanged. We refer the reader to \cite{angeris2020improved} for a general discussion of CFMM functionals as we will in the below focus on the particular instance of \textit{Geometric Mean Market Makers}.
	
	\medskip
	\noindent
	A Geometric Mean Market Maker (G3M) is a CFMM which is characterized by:
	
	\begin{itemize}
		\item a vector of weights $w\in \left(\mathbb{R}^+\right)^n$ such that $\Sigma_iw_i = 1$
		\item a liquidity providing fee $\phi \in [0,1[$
		\item a càdlàg inventory process $\theta_.: \mathbb{R}^+ \rightarrow \left(\mathbb{R}^+\right)^n$ with $\theta_i(t)$ the quantity of asset $S_i$ pooled within the G3M at time$~t$
		\item an exogenous trade process $\mathcal{T}$ such that:
		\begin{itemize}
			\item $\mathcal{T}_t$ is a collection of atomic trades faced by the G3M at time $t$: $\exists N_t \in \mathbb{N}~|~ \mathcal{T}_t = (\tau^k \in \mathbb{R}^n )_{k\le N_t}$
			\item Defining $\Delta_t :=  \theta_t -\theta_{t^-}$, we have $\Delta_t = \Sigma_k\tau^k$
		\end{itemize}
	\end{itemize}

	such that $\forall t \ge 0$, the G3M fundamental equation is satisfied:
	
	\begin{equation}\label{stateeq}
	\Pi_{i=1}^n\left(\theta^i_{t^-} + \Sigma_k\tau^k_i(1-\phi\mathbf{1}_{\tau^k_i>0})\right)^{w_i}= \Pi_{i=1}^n\left(\theta^i_{t^-}\right)^{w_i}.
	\end{equation}
	
	\bigskip
	
	Note that $\mathcal{T}_t$ is exogenous in the meaning that any agent in the market may post a trade against the G3M as long as (\ref{stateeq}) is satisfied.
	
	\section{No-arbitrage bounds}
	
	\subsection{G3M portfolio composition}
	
	\noindent
	Under the no-arbitrage hypothesis, there cannot exist a trade vector $\tau$ of strictly positive profit.
	
	\medskip
	
	Which implies that for $i \ne j, \tau_i > 0, -\theta_j < \tau_j < 0$

	\begin{eqnarray}
	 (\theta_i + (1-\phi)\tau_i)^{w_i}(\theta_j + \tau_j)^{w_j} = \theta_i^{w_i}\theta_j^{w_j} &\implies&  -\tau_iS_i - \tau_jS_j \le 0 \label{ineqproofarb_0}\\
	&\implies& \tau_i \ge -\tau_jS_j/S_i \label{ineqproofarb_1}\\
	&\implies&  (\theta_i - (1-\phi)\tau_jS_j/S_i)^{w_i}(\theta_j + \tau_j)^{w_j} \le \theta_i^{w_i}\theta_j^{w_j} \label{ineqproofarb}
	\end{eqnarray}
	
	\medskip
	
	Taking $\tau_j \rightarrow 0^-$,
	
		\begin{eqnarray}
	(\ref{ineqproofarb}) &\implies&   (1 - w_i(1-\phi)\tau_j \frac{S_j}{S_i\theta_i} + o(\tau_j))(1 + w_j\tau_j/\theta_j + o(\tau_j)) \le 1 \\
	&\implies&  -w_i(1-\phi)\tau_j \frac{S_j}{S_i\theta_i} + w_j\tau_j/\theta_j + o(\tau_j) \le 0 \\
	&\implies& \frac{\theta_i}{\theta_j} \ge (1-\phi)\frac{w_iS_j}{w_jS_i}
	\end{eqnarray}
	
	\medskip 
	
	Conversely, we obtain $ \frac{1}{1-\phi}\frac{w_iS_j}{w_jS_i} \ge \frac{\theta_i}{\theta_j}$ and conclude:

	\begin{equation}\label{noarb}
	\boxed{ (1-\phi)\frac{w_iS_j}{w_jS_i} \le \frac{\theta_i}{\theta_j} \le \frac{1}{1-\phi}\frac{w_iS_j}{w_jS_i}~~~~\forall (i,j) \in [[1,n]]^2}
	\end{equation}

	\subsection{G3M Wealth process}
	
	 \bigskip
	 
	 We define $K: t \mapsto \Sigma_i\theta^i_tS^i_t$ , the G3M wealth process which is representative of the market value of all assets pooled within the G3M.
	
	\begin{eqnarray}
	(\ref{noarb}) &\implies&(1-\phi)\frac{w_i\theta_k}{w_k}S_k \le \theta_iS_i \le \frac{1}{1-\phi}\frac{w_i\theta_k}{w_k}S_k  ~~~~~\forall i,k \\
	 &\implies&(1-\phi)\Sigma_{i}\frac{w_i\theta_k}{w_k}S_k \le K \le \frac{1}{1-\phi}\Sigma_{i}\frac{w_i\theta_k}{w_k}S_k ~~~~~\forall k \\
	 &\implies& (1-\phi)\frac{\theta_k}{w_k}S_k\le K \le \frac{1}{1-\phi}\frac{\theta_k}{w_k}S_k  ~~~~~\forall k\\
	 &\implies& (1-\phi)\Pi_k\frac{1}{w_k^{w_k}}\theta_k^{w_k}S_k^{w_k} \le K \le \frac{1}{1-\phi}\Pi_k\frac{1}{w_k^{w_k}}\theta_k^{w_k}S_k^{w_k} \label{arbvali}
	\end{eqnarray}

	\bigskip
	
	\noindent
	We define:
	
	\begin{itemize}
		\item $c := \Pi_i\frac{1}{w_i^{w_i}}$
		\item $k(t) := \Pi_i\theta_i^{w_i}$
	\end{itemize}

	and obtain:
	
	\begin{equation}
	(\ref{arbvali}) \implies \boxed{(1-\phi)ck(t)\Pi_iS_i^{w_i} \le K \le \frac{1}{1-\phi}ck(t)\Pi_iS_i^{w_i}}
	\end{equation}
		
	\bigskip
	
	\hphantom{~~~~}Note that if $\phi = 0$ then $k$ is constant and we have $K = ck\Pi_iS_i^{w_i}$.
	
	\bigskip
	
	\bigskip
	
	Which means that although the G3M essentially mirrors in composition a portfolio of fixed weights $(w_1,\dots,w_n)$, its dynamic is bounded by two geometrical-mean process times $k$, a functional representative of accumulated fees. The reader interested in the effect of transaction fees on the G3M tracking error to the constant-mix portfolio may refer to \cite{evans2021optimal}.
	
	\bigskip
	
	Let us now study how the G3M fares when compared to the portfolio of fixed weights $(w_1,\dots,w_n)$.
	
	\section{Impermanent Losses and G3M profitability}\label{secImpLoss}
	
	\subsection{Theoretical Derivation}
	
	\noindent
	We define $P$, the portfolio of fixed weights $(w_1,\dots,w_n)$, which is of dynamic:
	
	\begin{itemize}
		\item $P_0 = 1$
		\item $\frac{dP}{dt} = \langle w | \frac{dS}{S} \rangle$, $P_t = \exp\left(\int_0^t \left(\langle w | \mu \rangle  - \frac{1}{2}\langle w | \Sigma w \rangle\right) dt + \int_0^t\langle w | \sigma \odot dW_t\rangle \right)$
	\end{itemize}
	
	and $V := \Pi_iS_i^{w_i} = \exp\left(\int_0^t \left(\langle w | \mu \rangle  - \frac{1}{2}\langle w | \sigma^2 \rangle\right) dt + \int_0^t\langle w | \sigma \odot dW_t\rangle \right) $.
		
	\bigskip
	
	\noindent
	As such we obtain:
	
	\begin{equation}
	V  = P\exp\left(\frac{1}{2}\int_0^t\left(\langle w | \Sigma w \rangle -\langle w | \sigma^2 \rangle \right)dt\right)
	\end{equation}
	
	\medskip
	
	As $w \in \Omega := \{w \in \mathbb{R}^n | w \ge 0, \langle w | 1 \rangle =1 \}$ and $f: w \mapsto \langle w | \Sigma w \rangle -\langle w | \sigma^2 \rangle $ is a convex functional which is equal to zero on 
	$G := \{w \in \mathbb{R}^n | \exists j \in [[1,n]], w_i = \delta_{i,j} \forall i\}$, we recall that $\Omega$ is the convex closure of $G$ and obtain by direct application of the Jensen Theorem that $f \le 0$ on $\Omega$, i.e. that $V$ is doomed to underperform $P$. 
	
	\bigskip
	
	\noindent
	We define 
	
	\begin{equation}\label{ImpLossDef}
	\mathcal{I}_t := \exp\left(\frac{1}{2}\int_0^t\left(\langle w | \Sigma w \rangle -\langle w | \sigma^2 \rangle \right)dt\right)
	\end{equation} 
	
	the \textbf{Impermanent Loss} (IL) process of the G3M. Notice that its negative trend is essentially proportional to $\Sigma$ and this suggests that the G3M strategy is essentially `selling gamma' in option-trading language against the prospect of accruing transaction fees\footnote{See \cite{angeris2020does} and \cite{clark2020replicating} for the computation of CFMM greeks.}. As such, similarly to how traditional market makers widen their spread when volatility increases\cite{bollerslev1994bid,dayri2015large}, dynamically calibrating $\phi$ depending on volatility would likely allow G3M to optimise profitability in periods of market stress.
	
	Expanding on the comments of \cite{evans2020liquidity} on the underperformance of $V$ relative to the constant-mix portfolio $P$, and referring to the \textit{Stochastic Portfolio Theory} developed by Fernholz \cite{fernholz2009stochastic}, we further notice that $f$ is the direct opposite of the \textit{excess growth rate} provided by diversification in Fernholz' framework. In other words, investing in a G3M negates the boost provided by diversification in terms of portfolio growth rate. Indeed, one advantage of diversification is that the portfolio's growth rate is superior to the combination of its component stock growth: diversification in itself boosts portfolio returns. Alas, this is not the case for G3M strategies.

	Finally, note that $\Pi_iS_i^{w_i} \le \Sigma_iw_iS_i$, i.e. that  in the absence of trading fees, the G3M will \textit{always} underperform the buy-and-hold ``HODL'' portfolio\footnote{$\Sigma_i\log(S_i^{w_i}) = \Sigma_iw_i\log(S_i) \le \log(\Sigma_iw_iS_i)$} in addition to the constant-mix portfolio. 
	\subsection{Empirical study of Uniswap profitability for Liquidity provider}
	
	Having established the negative effect of Impermanent Loss on the G3M wealth process growth rate, we now study if transaction fees have \textit{empirically} been high enough to allow the G3M to outperform its constant-mix portfolio counterpart.
	
\begin{figure}[h]
		\begin{minipage}[c]{0.49\textwidth}
			\includegraphics[width=\linewidth]{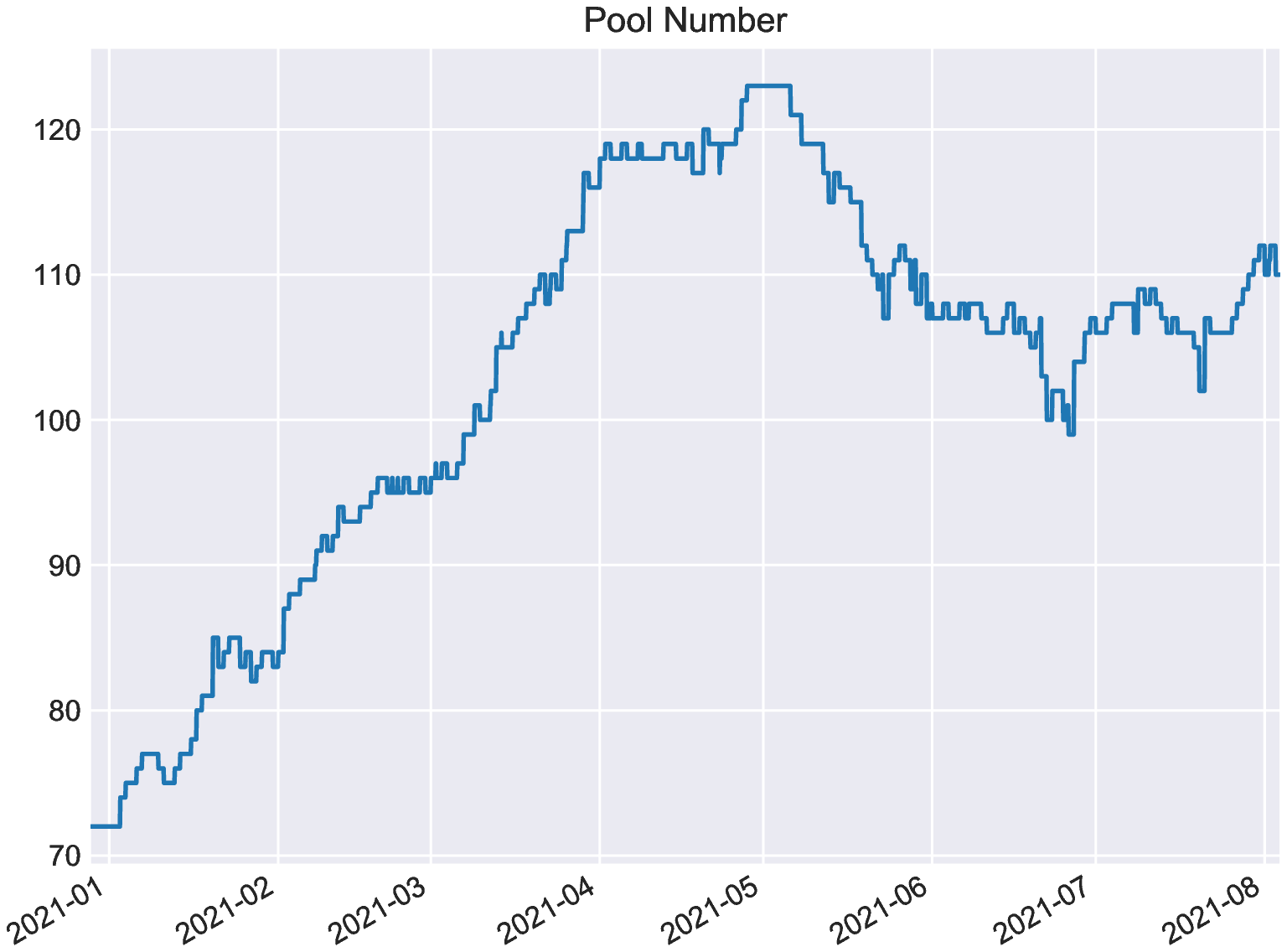}
			\caption{Number of UniswapV2 liquidity pool in the database.}
			\label{fig:univ2pools}
		\end{minipage}
		\hfill
		\begin{minipage}[c]{0.49\textwidth}
			\includegraphics[width=\linewidth]{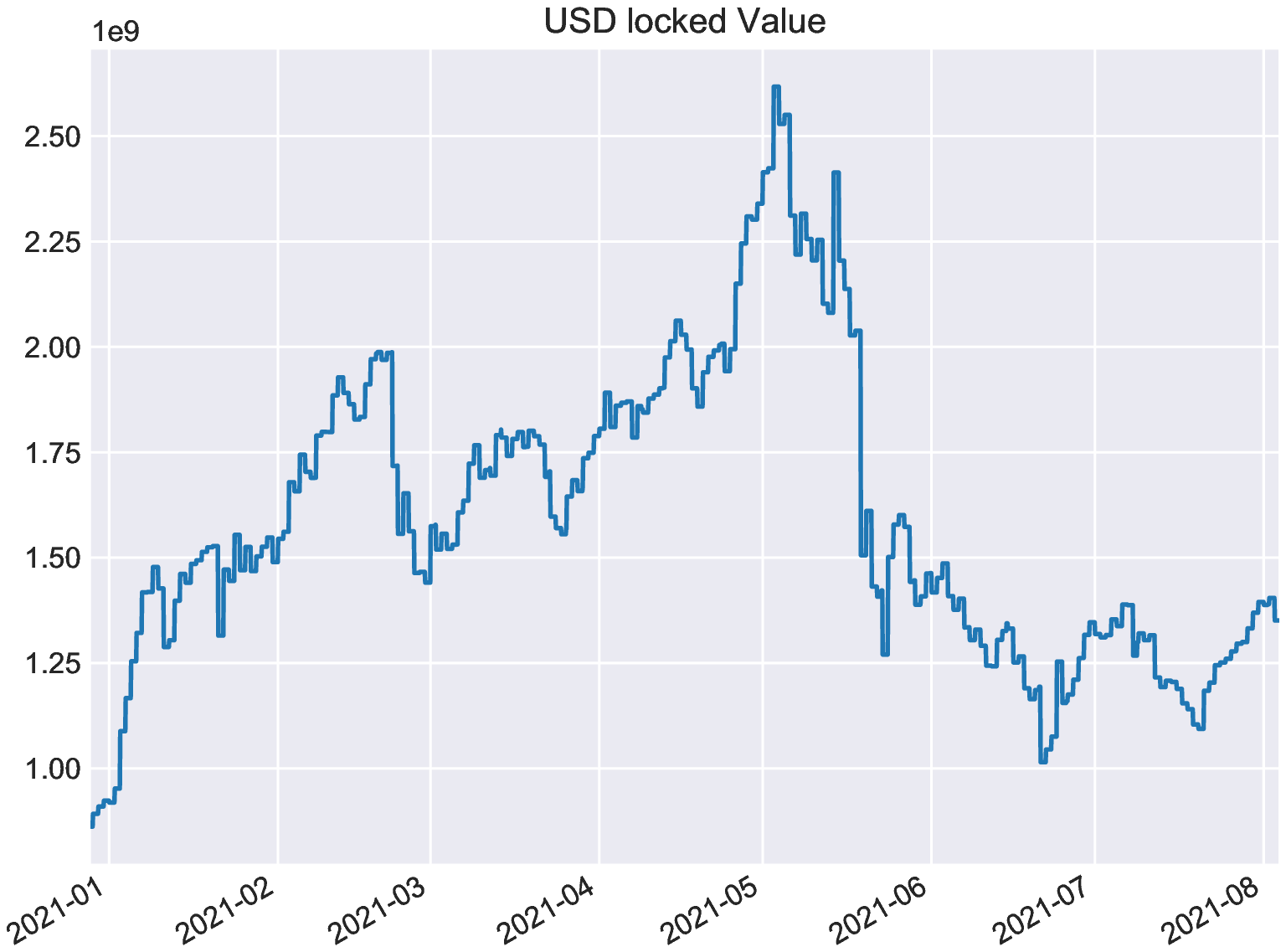}
			\caption{Total Value Locked in the database liquidity pools.}
			\label{fig:univ2TVL}
		\end{minipage}%
	\end{figure}

	To this effect, we studied a total of 147 UniswapV2 liquidity pools from 2020-12-28 to 2021-08-03 (\textit{Figure \ref{fig:univ2pools}} - 219 days, the data is sampled with an hourly frequency) with a liquidity superior to USD200'000 and totalling around \$1billion of locked value (\textit{Figure \ref{fig:univ2TVL}}). The data has been sourced from the PoolGenie analytics service\footnote{\url{www.poolgenie.xyz}}, which applies the above formula (\ref{ImpLossDef}) to compute Impermanent Loss and uses on-chain data for transaction fee metrics.

	\begin{figure}[h]
		\begin{minipage}[c]{0.49\textwidth}
			\includegraphics[width=\linewidth]{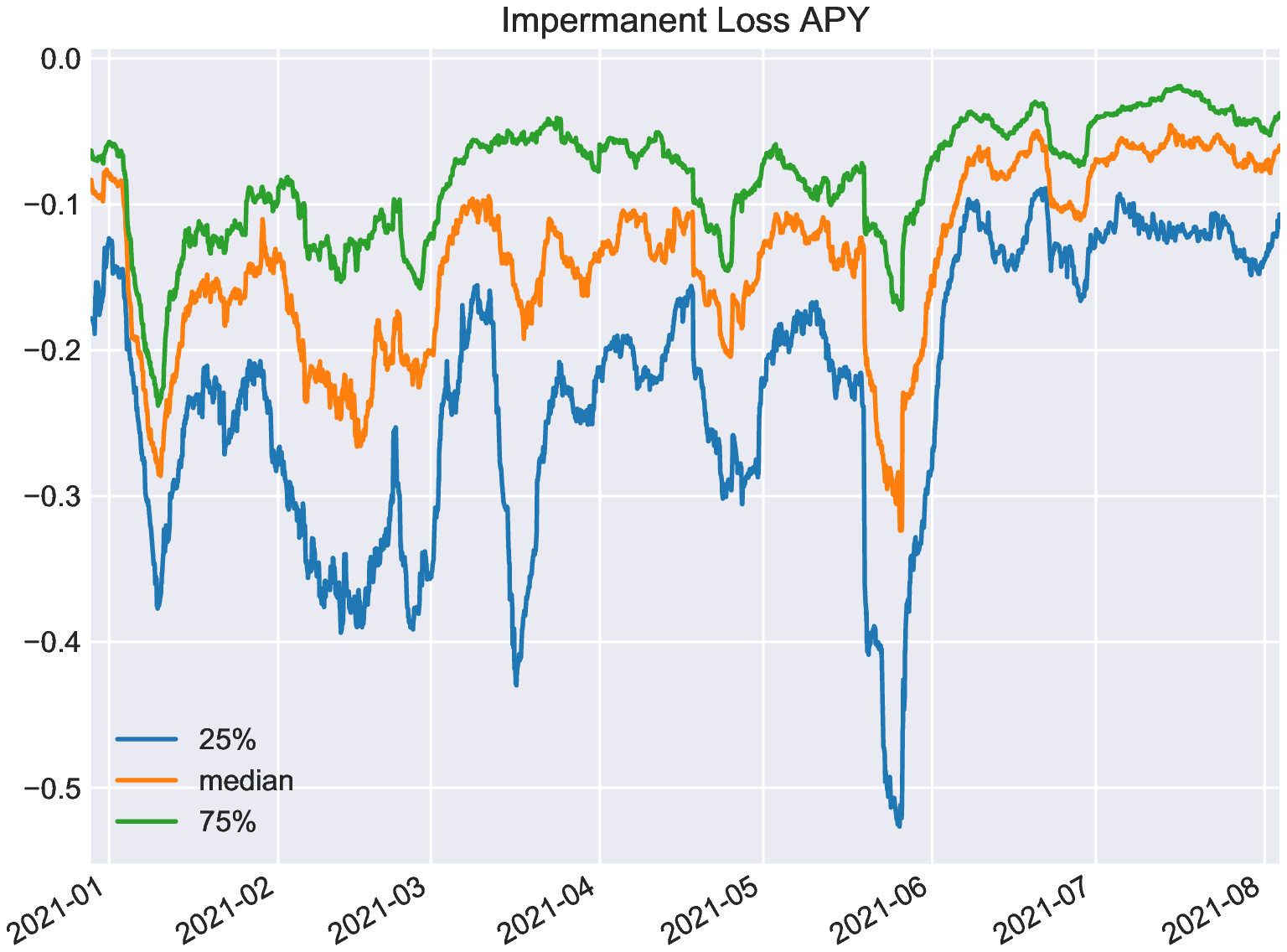}
			\caption{Uniswap pool Impermanent Loss APY quartiles.}
			\label{fig:IL}
		\end{minipage}
		\hfill
		\begin{minipage}[c]{0.49\textwidth}
			\includegraphics[width=\linewidth]{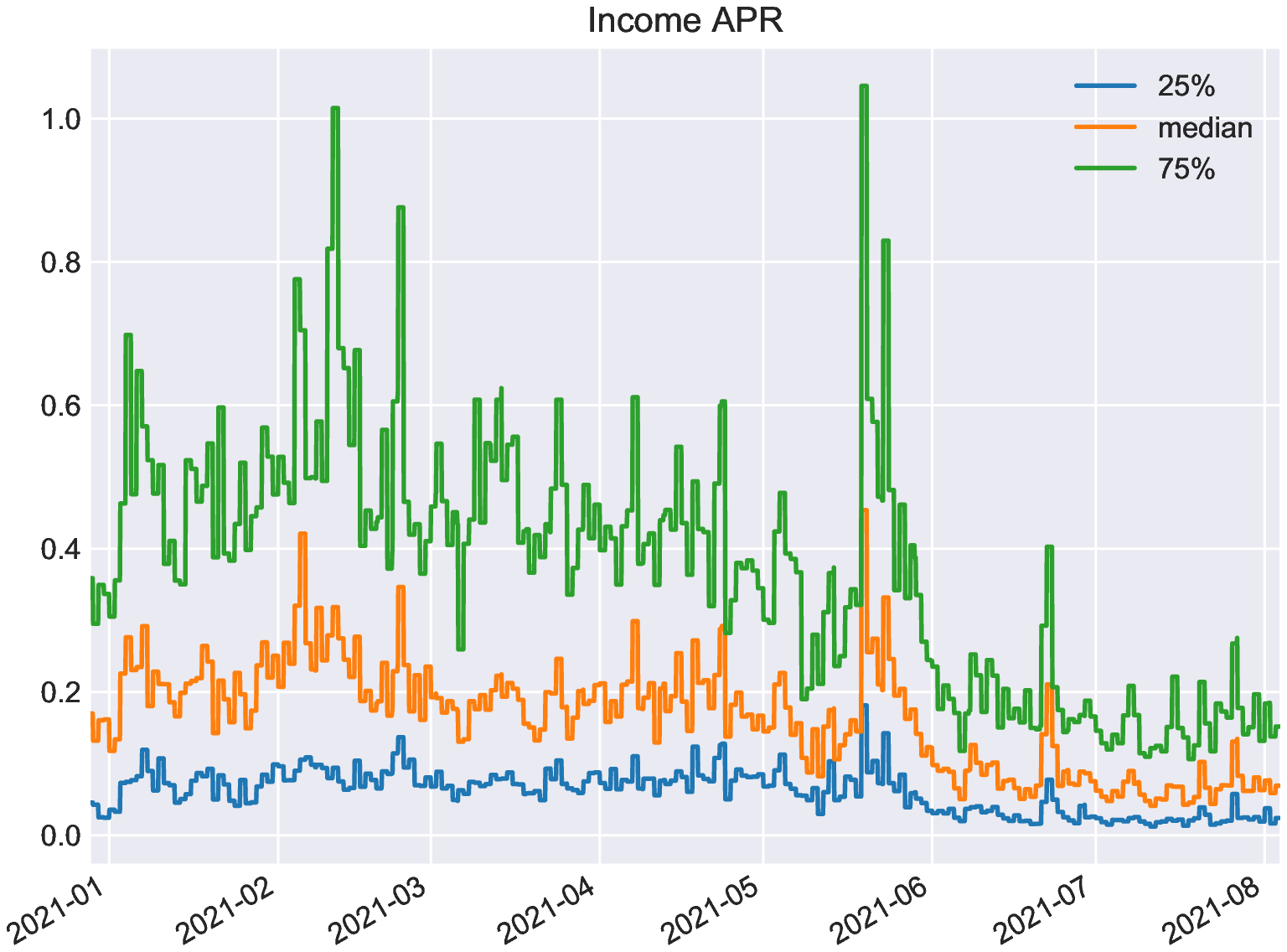}
			\caption{Uniswap pool income APR quartiles.}
			\label{fig:IROI}
		\end{minipage}%
	\end{figure}

\begin{figure}[h]
		\centering
		\includegraphics[width=.7\linewidth]{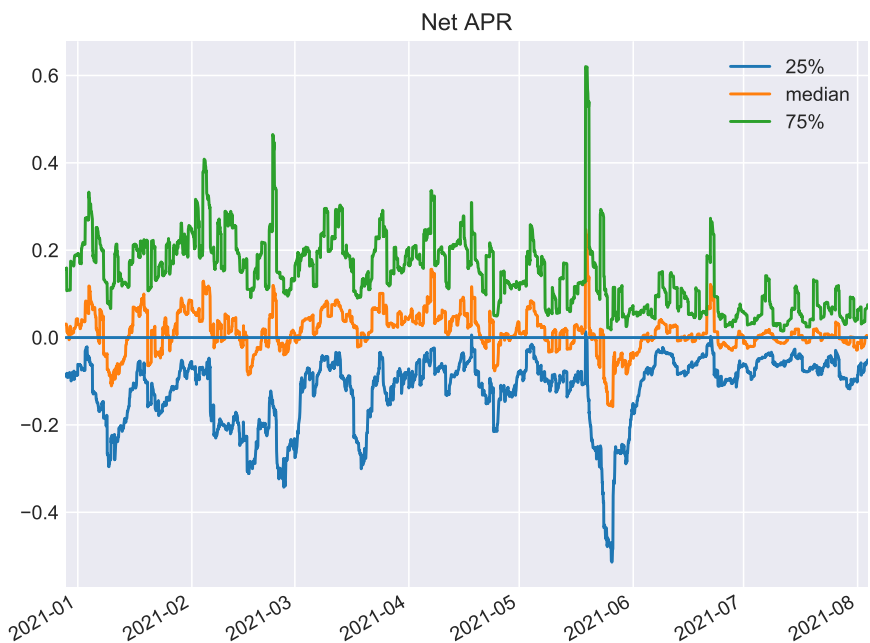}
		\caption{Uniswap pool net APR quartiles.}
		\label{fig:NROI}
	\end{figure}

We display in \textit{Figure \ref{fig:IL}} the different Impermanent Loss Annual Percentage Yield (APY) quartiles according to formula (\ref{ImpLossDef}). Consistent with intuition, given the very high volatility of crypto-assets, the Impermanent Loss trend is typically extremely high with many pools suffering from more than -20\% Impermanent Loss APY with impressive temporal as well as cross-sectional dispersions. In \textit{Figures \ref{fig:IROI}} and \textit{\ref{fig:NROI}}  are respectively displayed the Transaction fee Income Annual Percentage Rate (APR) as well as the total net G3M APR, which results from the addition of Impermanent Losses to accrued transaction fees.

\begin{figure}[h]
	\begin{minipage}[c]{0.49\textwidth}
		\includegraphics[width=\linewidth]{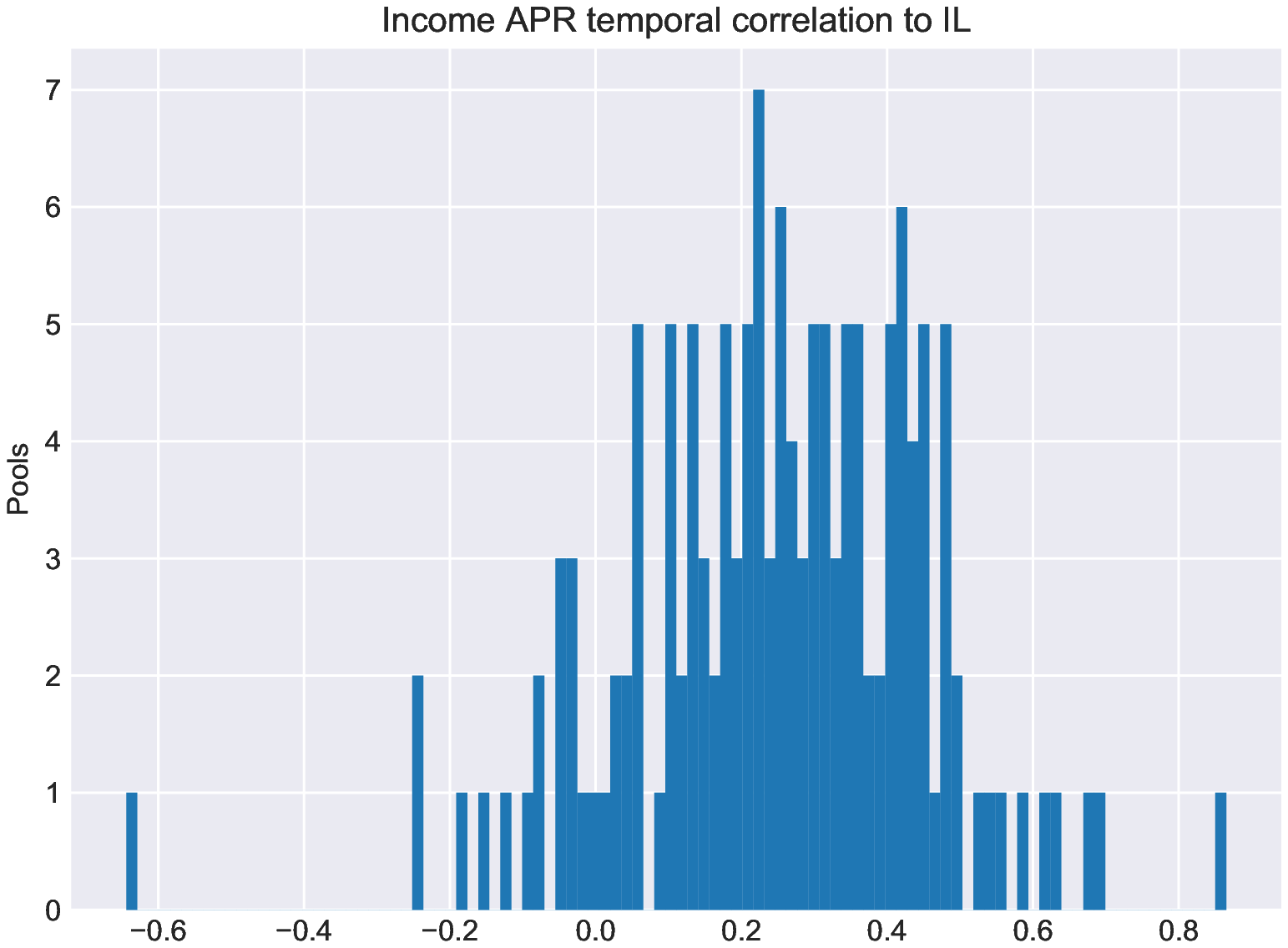}
		\caption{Uniswap pool transaction fee APR temporal correlation to IL histogram.}
		\label{fig:feeILcorrtemp}
	\end{minipage}
	\hfill
	\begin{minipage}[c]{0.49\textwidth}
		\includegraphics[width=\linewidth]{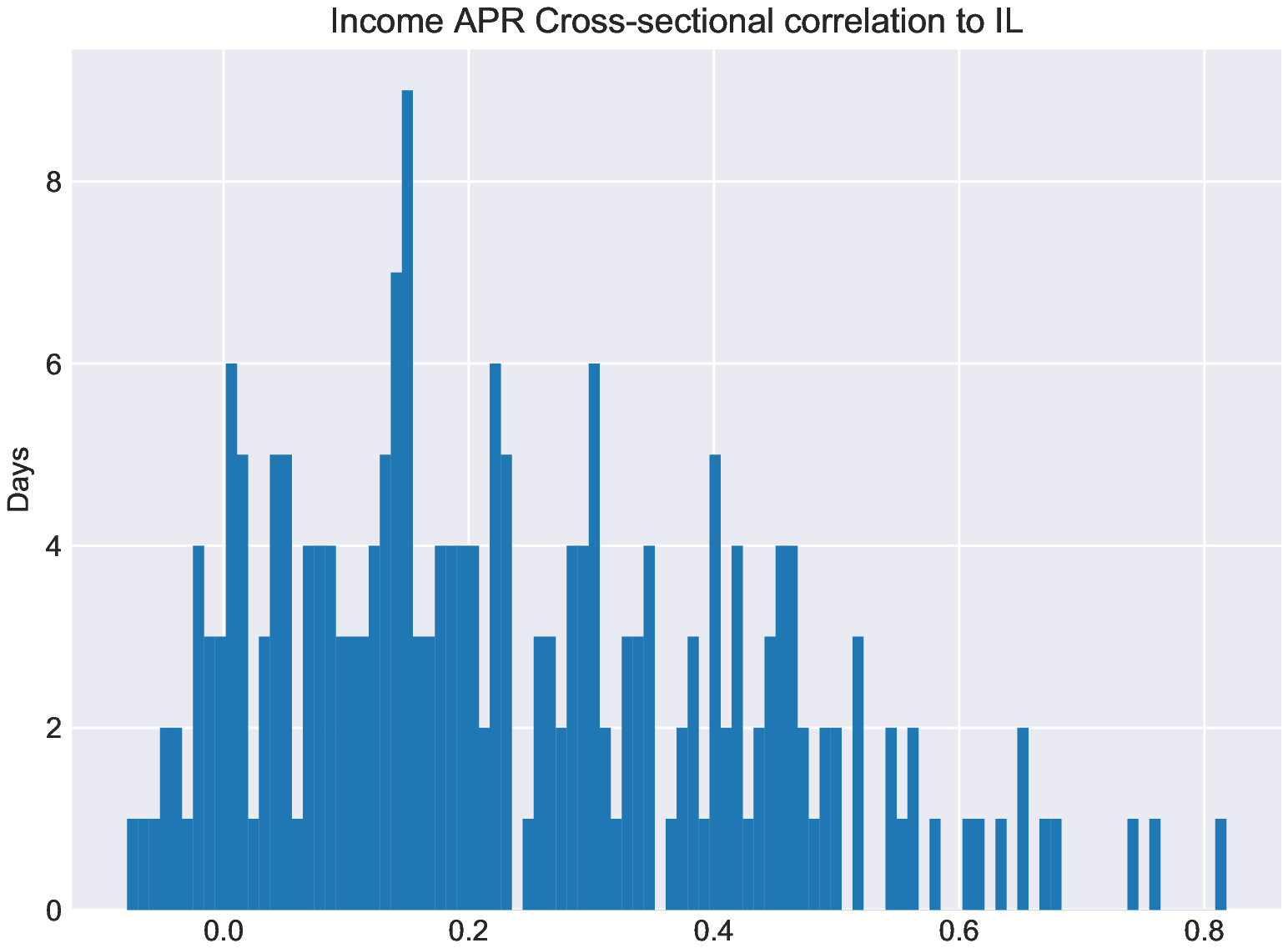}
		\caption{Uniswap pool transaction fee APR cross-sectional correlation to IL histogram.}
		\label{fig:feeILcorrtempX}
	\end{minipage}%
\end{figure}

The cross-sectional Income dispersion is also extreme with some pools yielding more than 50\% in transaction fees while others remain in the low single digits. Income APR is cross-temporally as well as cross-sectionally loosely correlated to the Impermanent Loss level (about 24\% on average for both cases) as more volatility comes with more trade volume and hence transaction fees (\textit{Figures \ref{fig:feeILcorrtemp}} and \textit{\ref{fig:feeILcorrtempX}}).

When adding everything up to obtain the net G3M APR (\textit{Figure \ref{fig:NROI}}), we witness that the median G3M pool has a low-to-nil trend when compared to the relevant constant-mix portfolio (median at 1\% p.a.). The average cross-sectional net APR interquartile range stands at 26\%.

\begin{figure}[h]
	\begin{minipage}[t]{0.49\textwidth}
		\includegraphics[width=\linewidth]{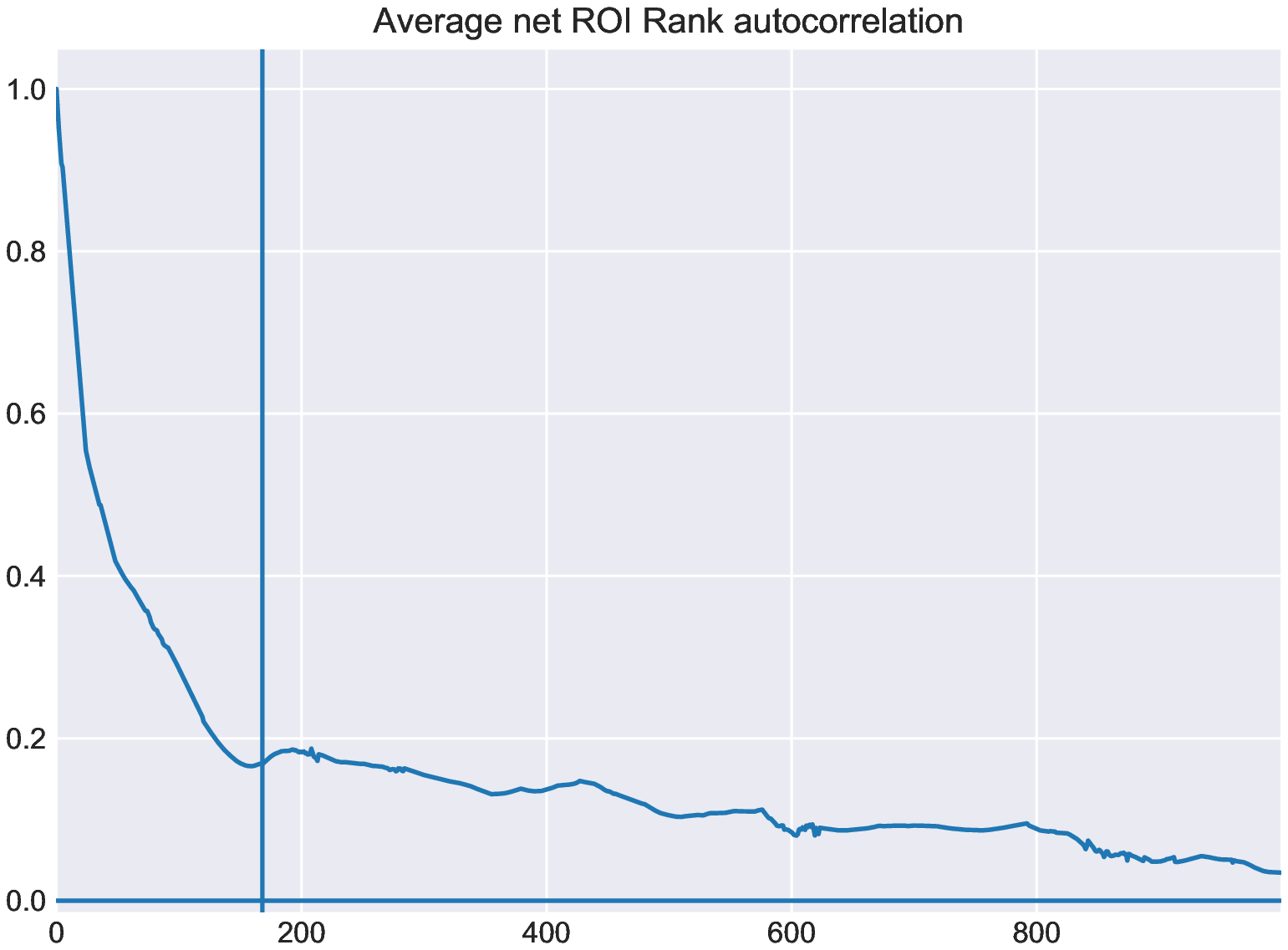}
		\caption{Net ROI autocorrelation (Spearman). Delta is in hours.}
		\label{fig:RanknROIautocorr}
	\end{minipage}
	\hfill
	\begin{minipage}[t]{0.49\textwidth}
		\includegraphics[width=\linewidth]{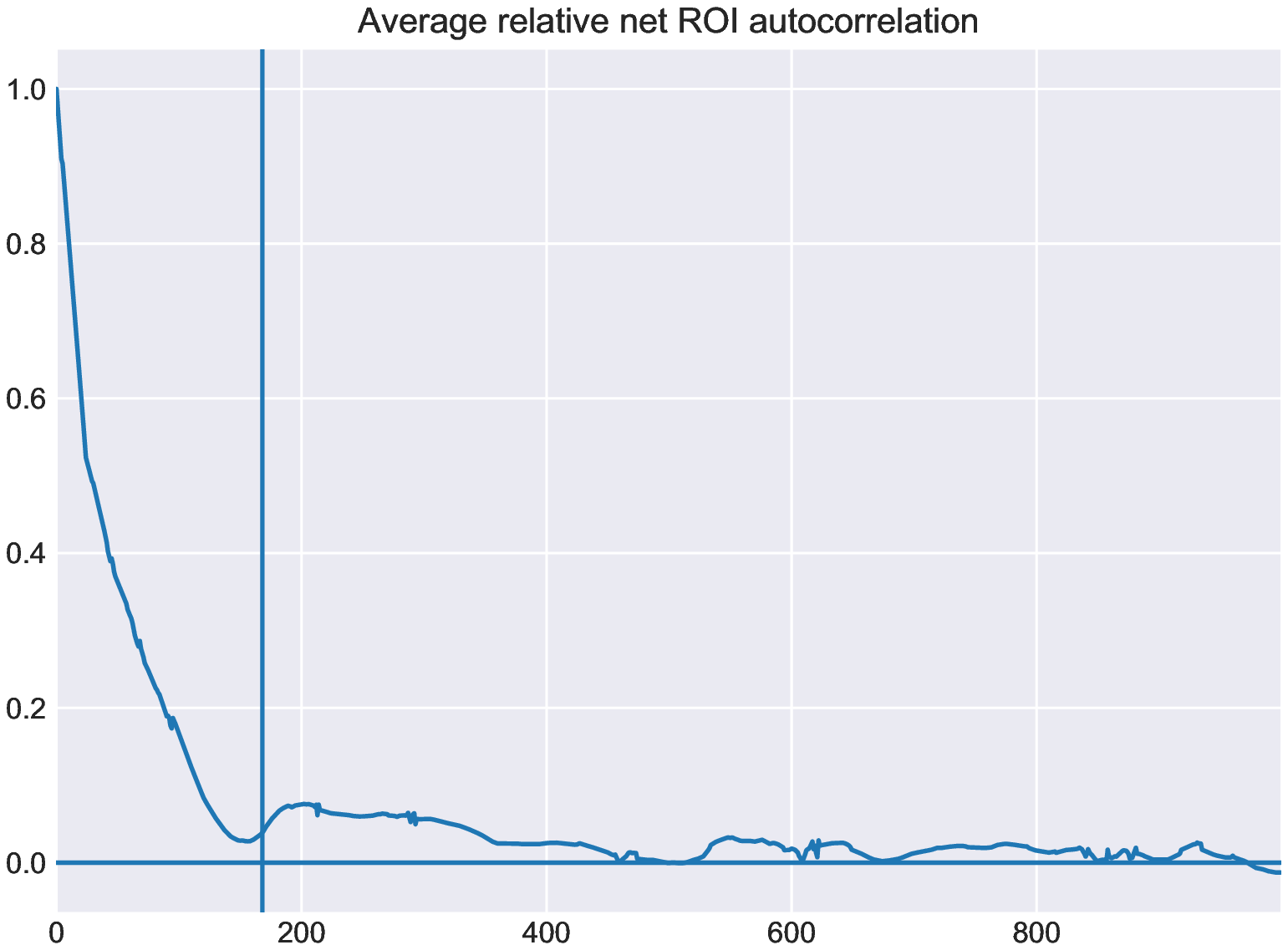}
		\caption{Relative Net ROI autocorrelation (Pearson). Where we centered each day the cross-sectional pool net ROI. Delta is in hours.}
		\label{fig:DeltanROIautocorr}
	\end{minipage}%
\end{figure}

We finally display the cross-sectional net ROI rank autocorrelation (\textit{Figure \ref{fig:RanknROIautocorr}}) as well as the average relative net ROI autocorrelation (\textit{Figure \ref{fig:DeltanROIautocorr}}), the latter being obtained by computing the Pearson autocorrelation of each pool's net ROI difference to the daily mean G3M net ROI. Given that we compute covariances using a weekly rolling window, we plotted a vertical line to mark the end of the computation-window. Although \textit{Figure \ref{fig:DeltanROIautocorr}} does not highlight any persistent autocorrelation in relative net ROI past a week's time mark, we can notice that the net ROI pool \textit{rank} remains well autocorrelated even after several weeks; this would imply that liquidity providers do not seamlessly switch pools in order to maximise their net ROI as the different liquidity pools would otherwise have more similar risk/return profiles. A cause of friction might be the unwillingness to change one's exposition to the various risky assets.

	\newpage
	
	\section{A note on UniswapV3}

	\subsection{Definition}
	
	In \cite{adams2021uniswap} Adams et al. introduce the innovation of range-bound or `concentrated' G3M liquidity pools as follows:
	
	\bigskip
	\noindent
	We Consider two assets $S^0,S^1$ and denote $p_t := S^0_t/S^1_t$, the price of $S^0$ expressed in the $S^1$ \textit{numerary}\footnote{similarly to \cite{adams2021uniswap}.} . The $(S^0,S^1)$ concentrated G3M of characteristics:
	
	\begin{itemize}
		\item respective inventories $x,y$ in $S^0,S^1$
		\item minimum price $p_a$, maximum price $p_b$
		\item total `Liquidity' $L$
	\end{itemize}

	distributes liquidity along the curve:
	
	\begin{equation}\label{univ3State}
	\left(x + \frac{L}{\sqrt{p_b}}\right)\left(y + L\sqrt{p_a}\right) = L^2
	\end{equation}
	
	while $p \in [p_a,p_b]$. Note that  for $\left(p_a=0, p_b = \infty\right)$ we are back to the traditional G3M state equation described above (\ref{stateeq}).
	
	\medskip
	
	We define:
	
	\begin{itemize}[leftmargin=1.5cm]
		\item $x^* := \left(x + \frac{L}{\sqrt{p_b}}\right)$
		\item $y^* := \left(y + L\sqrt{p_a}\right)$
	\end{itemize}

	the \textit{virtual reserves} of the concentrated G3M.

	\subsection{Leverage}
	
	While $p \in [p_a,p_b]$, we have the following identity:
	
	\begin{equation}
	(\ref{univ3State}) \implies    
	\begin{cases}
	px^* = y^* \\
	y^* = \sqrt{p}L
	\end{cases}  
	\implies    
	\begin{cases}
	x = L\left(1/\sqrt{p}-1/\sqrt{p_b}\right)\\
	y = L\left(\sqrt{p}-\sqrt{p_a}\right)
	\end{cases}  
	\end{equation}

	\newpage
	
	\noindent
	Denoting $K$ the G3M wealth process, we then write:
	
	\begin{eqnarray}
	K &=& (px+y)S_1 \\
	 &=& (2L\sqrt{p}-\frac{Lp}{\sqrt{p_b}}-L\sqrt{p_a})S_1\\
	K &=& 2LS_0^{1/2}S_1^{1/2} - \frac{L}{\sqrt{p_b}}S_0 - L\sqrt{p_a}S_1 \label{univ3Value}
	\end{eqnarray}
	
	\medskip
	Note that $L$ only depends on the quantity of tokens supplied and price at time $t_0=0$ as we have $L = y_{t_0}/(\sqrt{p_{t_0}}-\sqrt{p_a}) =  x_{t_0}\frac{\sqrt{p_bp_{t_0}}}{\sqrt{p_b}-\sqrt{p_{t_0}}}$.
	
	\medskip
	
	As such, the concentrated G3M strategy can be replicated by leveraging a traditional G3M position while simultaneously shorting the underlying tokens in order to finance the position.

	\bigskip
	\noindent
	Defining $\alpha = p/p_a>1$, $\beta = p_b/p\ge1$, we write:
	
	\begin{equation}
	(\ref{univ3Value}) \implies \boxed{K = 2LS_0^{1/2}S_1^{1/2}\left(1-\frac{1}{2}\left(\frac{1}{\sqrt{\alpha}}+\frac{1}{\sqrt{\beta}}\right)\right)}
	\end{equation}
	
	and define $l$, the \textit{leverage} of the concentrated G3M position as:
	
	\begin{equation}
	l :=\frac{1}{1-\frac{1}{2}\left(\frac{1}{\sqrt{\alpha}}+\frac{1}{\sqrt{\beta}}\right)} \ge 1
	\end{equation}
	
	so that $K = \frac{2LS_0^{1/2}S_1^{1/2}}{l}$

	\subsection{Impermanent Loss}
	
	Coming back to (\ref{univ3Value}), we have:
	
	\begin{eqnarray}
	\frac{dK}{K} &=& \frac{2LS_0^{1/2}S_1^{1/2}}{K}\frac{d\left(S_0^{1/2}S_1^{1/2}\right)}{S_0^{1/2}S_1^{1/2}} - \frac{LS_0}{\sqrt{p_b}K}\frac{dS_0}{S_0} - \frac{L\sqrt{p_a}S_1}{K}\frac{dS_1}{S_1} \\
	 &=& l\left(\frac{d\left(S_0^{1/2}S_1^{1/2}\right)}{S_0^{1/2}S_1^{1/2}} - \frac{1}{2\sqrt{\beta}}\frac{dS_0}{S_0}  -\frac{1}{2\sqrt{\alpha}}\frac{dS_1}{S_1}\right) \\
	 &=& l\left( \left(\frac{1}{2}\frac{dS_0}{S_0} + \frac{1}{2}\frac{dS_1}{S_1} + \left(\frac{\sigma_0^2}{4}+ \frac{\sigma_1^2}{4} + \frac{\rho\sigma_0\sigma_1}{2} -  \frac{\sigma_0^2}{2} - \frac{\sigma_1^2}{2}\right)dt\right)  - \frac{1}{2\sqrt{\beta}}\frac{dS_0}{S_0}  -\frac{1}{2\sqrt{\alpha}}\frac{dS_1}{S_1}\right) \\
	 	 &=& \frac{l}{2}\left(\left(1-\frac{1}{\sqrt{\beta}}\right)\frac{dS_0}{S_0} + \left(1-\frac{1}{\sqrt{\alpha}}\right)\frac{dS_1}{S_1}  -\left(\frac{\sigma_0^2}{2}+ \frac{\sigma_1^2}{2} - \rho\sigma_0\sigma_1\right)dt \right). \label{univ3dynamic}
	\end{eqnarray}
	
	\medskip
	
	Where we see that the concentrated G3M Impermanent Loss dynamics are essentially the same as in section \ref{secImpLoss} although the relative exposure (`delta') to the pool components differs.
	
	\medskip
	
	In cases where $\lambda := \alpha = \beta$, i.e. the investor chooses to dynamically rebalance her position so as to always have the current price be at the centre of her liquidity providing interval, (\ref{univ3dynamic}) simplifies to:
	
	\begin{equation}\label{univ3Centered}
	\frac{dK}{K} = \frac{1}{2}\frac{dS_0}{S_0} + \frac{1}{2}\frac{dS_1}{S_1} - \frac{\sqrt{\lambda}}{\sqrt{\lambda}-1}\left(\frac{\sigma_0^2}{2}+ \frac{\sigma_1^2}{2} - \rho\sigma_0\sigma_1\right)dt
	\end{equation}
	
	where we assume rebalancing is done with no friction.
	
	\medskip
	
	\noindent
	
	In such cases, concentrating the liquidity on smaller intervals directly amounts to leveraging the Impermanent Loss/Transaction Fees component of the standard G3M wealth process. The obtained position can indeed be replicated by leveraging a UniswapV2 pool while simultaneously hedging part of the token price exposition by shorting the relevant constant-mix portfolio. Using a non-centered interval essentially amounts to unequally hedging the two liquidity pools tokens.

	\section{Conclusion}
	
	When compared to constant-mix portfolios, investing in a G3M comes down to forsaking the benefits of portfolio diversification in terms of growth rate for the prospect of earning transaction fees.
	
	As a consequence, in the absence of transaction fees, these AMMs will always underperform their constant-mix portfolio counterparts as well as the unrebalanced `HODL' alternative. 
	
	Empirically, transaction fee income has in median just been high enough to compensate Impermanent Losses; furthermore the dispersion of net income  has been significant as the cross-sectional net APR interquartile range stood at 26\% p.a. in our dataset. Past data however suggests that informed liquidity providers may know in advance which pools are more likely to provide a positive investment return.
	
	The above results also apply to UniswapV3 `concentrated range' liquidity positions as they are essentially equivalent to leveraging the Transaction Fee/Impermanent Loss trade-off of standard G3Ms.
	
	\medskip
	Caution is hence warranted when considering to use a G3M setting \textit{in lieu} of a traditionally managed constant-mix portfolio. As alluring as it may appear to ``collect fees from traders, who rebalance your portfolio by following arbitrage opportunities'' instead of ``paying fees to portfolio managers''\cite{Martinelli2019Balancer}, this is not a free-lunch and the success of the operation hinges on the prospect of levying enough transaction fees from non-arbitrageurs to offset the inevitable Impermanent Losses.

	\medskip
	
	Further research could point to modelling the transaction-fee component of G3M returns and studying the cumulative fee process as a function of the percentage of `non-informed' vs `informed' traders. The Impermanent Loss dynamics of other AMM designs, such as Curve \cite{egorov2021curve}, could also be of interest.

	\bibliographystyle{unsrt}
	\bibliography{ImpLossBib}
	
\end{document}